\documentclass[aps,prd,nofootinbib,twocolumn,groupedaddress]{revtex4}
\usepackage{dcolumn}
\usepackage{bm}
\usepackage{pdfpages}
\usepackage{amsmath}    
\usepackage{amsfonts} 
\usepackage{amssymb}
\usepackage{mathptmx}      
\usepackage{graphicx}   
\usepackage{mathtools}
\usepackage{graphics}
\usepackage{epsfig}
\usepackage{tabularx}   
\usepackage{multirow}
\usepackage{bibentry}
\usepackage{braket}
\usepackage{algorithm}
\usepackage{algorithmic}
\usepackage{listings}
\usepackage{setspace}
\usepackage{fancyhdr}
\usepackage{slashed}
\usepackage{color}
\usepackage{xfrac}
\def\Journal#1#2#3#4{{#1} {\bf#2}, #3 (#4)}

\def\NPA{{\rm Nucl. Phys.} A}
\def\NPB{{\rm Nucl. Phys.} B}
\def\PLB{{\rm Phys. Lett.}  B}

\def\PRD{{\rm Phys. Rev.} D}
\def\PRC{{\rm Phys. Rev.} C}

\def\ep{\epsilon}
\def\vep{\varepsilon}

\def\ra{\rangle}

\def\lam{\lambda}

\def\be{\begin{equation}}
\def\ee{\end{equation}}
\def\bea{\begin{eqnarray}}
\def\eea{\end{eqnarray}}
\begin{document}
\title{Doubly virtual $(\pi^0,\eta,\eta')\to\gamma^*\gamma^*$ transition form factors in the light-front quark model}
\author{ Ho-Meoyng Choi\\
{\em Department of Physics, Teachers College, Kyungpook National University,
     Daegu, Korea 41566}\\
     Hui-Young Ryu\\
{\em Department of Physics, Pusan National University,
     Pusan, Korea 46241}\\         Chueng-Ryong Ji\\
{\em Department of Physics, North Carolina State University,
Raleigh, North Carolina 27695-8202, USA} }
\begin{abstract}
We report our investigation on the doubly  virtual transition form factors (TFFs) $F_{{\rm P}\gamma^*}(Q^2_1,Q^2_2)$ for the 
${\rm P}\to\gamma^*(q_1)\gamma^*(q_2) \;({\rm P}=\pi^0,\eta,\eta')$ transitions  using the light-front quark model (LFQM).  
Performing a LF calculation in the exactly solvable manifestly covariant Bethe-Salpeter (BS) model as the first illustration, 
we use the $q^+_1=0$ frame and find that both LF and manifestly covariant calculations produce
exactly the same results for $F_{{\rm P}\gamma^*}(Q^2_1,Q^2_2)$. This confirms the absence of the LF zero mode in the doubly 
virtual TFFs. We then map  this covariant BS model to the standard LFQM using the more phenomenologically accessible 
Gaussian wave function provided by the LFQM analysis 
of meson mass spectra.
For the numerical analyses of $F_{{\rm P}\gamma^*}(Q^2_1,Q^2_2)$, 
we compare our LFQM results with the available experimental data and the perturbative QCD (pQCD) and vector meson dominance
(VMD) model predictions.
As $(Q^2_1, Q^2_2)\to\infty$, our LFQM result for doubly virtual TFF is consistent with the pQCD prediction, 
i.e. $F_{{\rm P}\gamma^*}(Q^2_1, Q^2_2)\sim 1/(Q^2_1 + Q^2_2)$, while it differs greatly from 
the result of the VMD model, which behaves as $F^{\rm VMD}_{{\rm P}\gamma^*}(Q^2_1, Q^2_2)\sim 1/(Q^2_1 Q^2_2)$. Our LFQM prediction for 
$F_{\eta'\gamma^*}(Q^2_1,Q^2_2)$ shows an agreement with the very recent experimental data  obtained from the $BABAR$ Collaboration
for the ranges of $2< (Q^2_1, Q^2_2) <60$ GeV$^2$.
\end{abstract}
\maketitle
\section{Introduction}
The meson-photon transitions such as ${\rm P}\to\gamma^{(*)}\gamma^{(*)}({\rm P}=\pi^0,\eta,\eta')$
with one or two virtual photons have been of interest to both theoretical and experimental physics
communities since they are the simplest possible bound state processes in quantum chromodynamics (QCD) and
they play a significant role in allowing both the low- and high-energy precision tests of the standard model.

In particular, both singly  virtual and doubly virtual transition form factors (TFFs) are required to estimate
the hadronic light-by-light (HLbL) scattering contribution to the muon anomalous magnetic moment $(g-2)_\mu$. 
The HLbL contribution is in principle obtained by integrating some weighting functions times the product of a single-virtual and
a double-virtual TFF for  spacelike momentum~\cite{JN,Ny2016,Lattice16}. 
The single-virtual TFFs have been measured either from the spacelike $e^+e^-\to e^+e^- {\rm P}$ process in the single tag 
mode~\cite{CELLO91,CLEO98,BES15_Pi} or from the timelike Dalitz decays 
${\rm P}\to\bar{\ell}\ell\gamma$~\cite{NA60,NA60-17,A22014, A22011,A2pi,BES15}
where $(2m_\ell)^2\leq q^2\leq m^2_{\rm P}$. The timelike region beyond the single Dalitz decays may be accessed 
through the $e^+e^- \to {\rm P}\gamma$ annihilation processes, and 
the $BABAR$ Collaboration~\cite{BABAR06} measured the timelike $F_{\eta^{(\prime)}\gamma}$ TFFs
from the reaction $e^+e^-\to\eta^{(\prime)}\gamma$ at an average $e^+e^-$ center of mass energy of
$\sqrt{s}=10.58$ GeV. 

\begin{figure}
\begin{center}
\includegraphics[height=3.5cm, width=6cm]{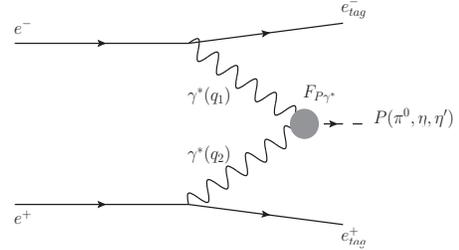}
\caption{\label{fig1} The diagram for the $e^+e^-\to e^+e^-{\rm P}$ process. }
\end{center}
\end{figure}
Very recently, the $BABAR$ Collaboration~\cite{BABAR18} measured for the first time the double-virtual  
$\gamma^*(q_1)\gamma^*(q_2)\to\eta'$ TFF $F_{\eta'\gamma^*}(Q^1_1,Q^2_2)$ 
in the spacelike(i.e. $Q^2_{1(2)}=-q^2_{1(2)}>0$) kinematic region
of $2<Q^2_1, Q^2_2<60$ GeV$^2$ by using the $e^+e^-\to e^+e^-\eta'$ process in the double-tag mode as shown
in Fig.~\ref{fig1}.
It is very interesting to note that the measurement of $F_{{\rm P}\gamma^*}(Q^1_1,Q^2_2)$ at large $Q^2_1$
and $Q^2_2$ distinguishes the predictions of the model inspired by perturbative QCD(pQCD)~\cite{BL80,Braaten83},
$F^{pQCD}_{{\rm P}\gamma^*}(Q^2_1, Q^2_2)\sim 1/(Q^2_1 + Q^2_2)$, from those of the vector meson
dominance (VMD) model~\cite{VDM1,VDM2,VDM3}, $F^{VDM}_{{\rm P}\gamma^*}(Q^2_1, Q^2_2)\sim 1/(Q^2_1 Q^2_2)$, while
both models predict the same asymptotic dependence $F^{asy}_{{\rm P}\gamma}(Q^2,0)\sim 1/Q^2$ as $Q^2\to\infty$.

The low-energy behavior of the TFF for the doubly virtual $\pi^0\to\gamma^*\gamma^*$ transition was recently 
investigated within a Dyson-Schwinger and Bethe-Salpeter~(BS) framework~\cite{Weil}.
In our previous analysis~\cite{CRJ17}, we explored the TFF $F_{P\gamma}(Q^2,0)$
for the single-virtual ${\rm P}\to\gamma^*\gamma ~({\rm P}=\pi^0,\eta,\eta')$ transition both in the spacelike and timelike
region using the light-front quark model (LFQM)~\cite{CJ_99,CJ_DA,PiGam16,CJ_PLB,CJBc}.
In particular, we presented the new direct method to explore the timelike region without resorting to mere analytic continuation
from a spacelike to a timelike region. Our direct calculation in the timelike region has shown the complete agreement with 
not only the analytic continuation result from the spacelike region but also the result from the dispersion relation between
the real and imaginary parts of the form factor.

The purpose of this work is to extend our previous analysis~\cite{CRJ17} to compute the TFF for the doubly virtual 
$\eta'\to\gamma^*\gamma^*$  transition and compare with the recent $BABAR$ data for 
$F_{{\rm P}\gamma^*}(Q^2_1,Q^2_2)$~\cite{BABAR18}. We also present the TFFs for $(\eta, \pi^0)\to\gamma^*\gamma^*$
as well to complete the analysis of doubly virtual photon-pseudoscalar meson transitions in our LFQM.

The paper is organized as follows. In Sec.~\ref{sec:II}, we discuss the
TFFs for the doubly virtual ${\rm P}\to\gamma^*\gamma^*$ transitions  in an exactly solvable model first based on the covariant 
BS model of
(3+1)-dimensional fermion field theory
to check the existence (or
absence) of the LF zero mode~\cite{Zero1,Zero2,Zero3,Zero4}
as one can pin down the zero mode exactly in the manifestly covariant 
BS model~\cite{BCJ02,BCJ03,TWV, TWPS,TWPS17}. 
Performing both the manifestly covariant
calculation and the LF calculation, we explicitly show the equivalence between the
two results and the absence of the zero-mode contribution to the TFF. 
The $\eta-\eta'$ mixing scheme for the calculations of the $(\eta,\eta')\to\gamma^*\gamma^*$
TFFs is also introduced in this section.
In Sec.~\ref{sec:III}, we apply the self-consistent correspondence relations~[see, e.g., Eq. (35) in~\cite{TWPS}~] between 
the covariant BS model and the LFQM and we present the standard LFQM calculation with 
the more phenomenologically accessible model wave functions 
provided by the LFQM analysis of meson mass spectra~\cite{CJ_PLB,CJ_99}.
In Sec.~\ref{sec:IV}, we present our numerical results for the
$(\pi^0,\eta,\eta')\to\gamma^*\gamma^*$ TFFs
and compare them with the available experimental data. Summary and discussion follow in Sec.~\ref{sec:V}. 

\section{Manifestly Covariant Model}
\label{sec:II}
\begin{figure}
\begin{center}
\includegraphics[height=2.cm, width=8cm]{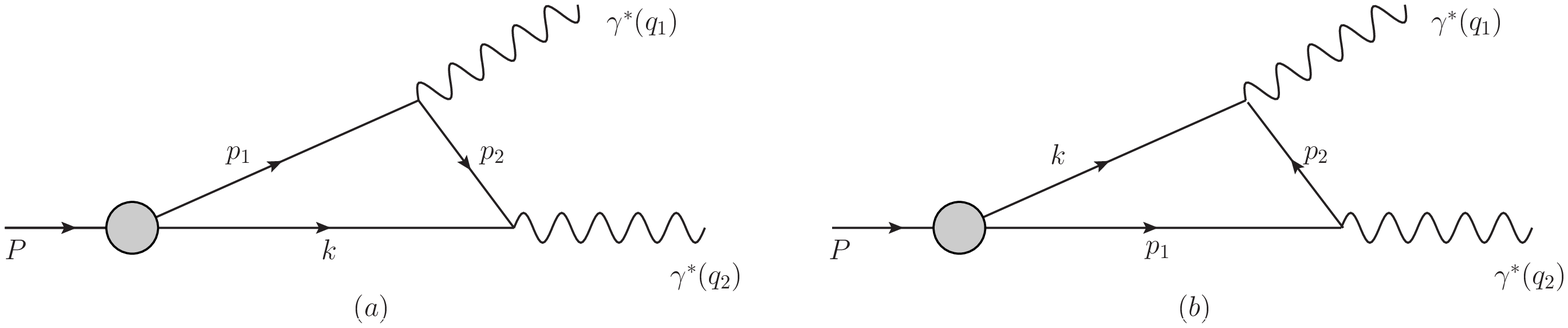}
\caption{\label{fig2} One-loop Feynman diagrams that contribute to ${\rm P}\to\gamma^*\gamma^*$.
(a) and (b)  represent the amplitudes of the virtual photon with momentum $q_1$ being attached to the
quark and antiquark lines. }
\end{center}
\end{figure}
The TFF $F_{{\rm P}\gamma^*}$ for 
the doubly virtual ${\rm P}(P)\to\gamma^*(q_1)\gamma^*(q_2)$~(${\rm P}=\pi^0, \eta, \eta'$) transition is defined via the amplitude 
$T$
as follows:
\be\label{Eq1}
T = i e^2 F_{{\rm P}\gamma^*}(q^2_1,q^2_2)\ep^{\mu\nu\rho\sigma}\ep_{1\mu}\ep_{2\nu}q_{1\rho}q_{2\sigma},
\ee
where $P$ is the four-momenta of the pseudoscalar meson, 
$q_{1(2)}$ and $\vep_{1(2)}$ are the momenta and polarization vectors of  two virtual photons 1 and 2,
respectively. This process is illustrated by the one-loop Feynman diagrams in Figs.~\ref{fig2}(a) and~\ref{fig2}(b),
which represent the amplitudes of the virtual photon with momenta $q_1$ being attached to the
quark and antiquark lines, respectively.  While we shall only discuss the amplitude shown in Fig.~\ref{fig2}(a),
the total amplitude should of course include the contribution from the process in Fig.~\ref{fig2}(b) as well.

In the exactly solvable manifestly covariant BS model, the covariant amplitude $T$ in Fig.~\ref{fig2} (a) 
is obtained with the following momentum integral:
\be\label{Eq2}
T = i e_Q e_{\bar Q} N_c
\int\frac{d^4k}{(2\pi)^4} \frac{H_0}
{N_{p_1} N_k N_{p_2}}S,
\ee
where $N_c$ is the number of colors and $e_Q~(e_{\bar Q})$ is the quark~(antiquark) electric charge.
The denominators $N_{p_j} (= p_{j}^2 -m_Q^2 +i\ep)~ (j=1,2)$
and $N_k(= k^2 - m_{\bar Q}^2 + i\ep)$ come from the intermediate quark and antiquark propagators
of mass $m_Q=m_{\bar Q}$ carrying the internal
four-momenta $p_1=P-k$, $p_2=P-q-k$, and $k$, respectively.
The trace term $S$ in Eq.~(\ref{Eq2}) is obtained as
\bea\label{Eq3}
 S &=&  {\rm Tr}\left[\gamma_5\left(\slash \!\!\!\!\!p_1 + m_Q \right)
 \slash \!\!\!\!\!\ep_1 \left(\slash \!\!\!\!\!p_2 + m_Q \right)\slash \!\!\!\!\!\ep_2
 \left(-\slash \!\!\!\!k + m_Q \right) \right]
 \nonumber\\
&=& 4 i m_Q \ep^{\mu\nu\rho\sigma}\ep_{1\mu}\ep_{2\nu}q_{1\rho}q_{2\sigma}.
\eea
For the ${\bar q}q$ bound-state vertex function $H_0=H_0 (p^2_1, k^2)$ of the
meson, we simply take the constant parameter $g$ in our model calculation.
The covariant loop is regularized properly with this constant vertex. 

Using the Feynman parametrization for the three propagators $1/(N_{p_1} N_{k} N_{p_2})$,
we obtain the manifestly covariant result by defining the amplitude in Fig.~\ref{fig1}(a) as
$T_{(a)} = ie_Q e_{\bar Q} [I^{m_Q}_{(a)}]^{\rm Cov}(q^2_1, q^2_2)\ep^{\mu\nu\rho\sigma}\ep_{1\mu}\ep_{2\nu}q_{1\rho}q_{2\sigma}$,
where
\bea\label{Eq6}
[I^{m_Q}_{(a)}]^{\rm Cov} &=& \frac{N_c g}{4\pi^2} \int^1_0 dx\int^{1-x}_0 dy
\nonumber\\
&&\times
\frac{m_Q}{(x+y-1)(x M^2 - y Q^2_2 ) + xy Q^2_1 + m^2_Q},
\eea
with the physical meson mass $M$. 

For the LF calculation in parallel with the manifestly covariant one,
we use  the  $q^+_1=0$ frame, where we take $P=( P^+, M^2/P^+, 0)$;
$q_1=(0, q^-_1, {\bf q}_{1\perp})$; and $q_2=(P^+, (q^2_2 +{\bf q}^2_{1\perp})/P^+, -{\bf q}_{1\perp})$
 so that  $q^2_1 =-{\bf q}^2_{1\perp} \equiv -Q^2_1$ and $q^2_2 = -Q^2_2$.

In this frame,  the Cauchy integration of Eq.~(\ref{Eq2}) over $k^-$ in Fig.~\ref{fig2}(a) yields
\be\label{Eq8}
 [I^{m_Q}_{(a)}]^{\rm LF} = \frac{N_c }{4\pi^3}\int^{1}_0
 \frac{dx}{x(1-x)} \int d^2{\bf k}_\perp
 \frac{m_Q}{ Q^2_2 + M^{\prime 2}_0} \chi(x,{\bf k}_\perp),
\ee
where $x$ is the LF longitudinal  momentum fraction defined by $k^+=(1-x)P^+$
and the LF $({\rm P}q\bar{q})$-vertex function 
\be\label{NEq9}
\chi(x,{\bf k}_\perp) = \frac{g}{x (M^2 -M^2_0)} 
\ee
is the ordinary LF valence wave function with  $M^2_0 = \frac{ {\bf k}^{2}_\perp + m^2_Q}{x (1-x)}$ being the invariant mass. 
Note here that the pole of $N_k = 0$ is taken for the Cauchy integration to get Eq.(\ref{NEq9}).
The primed momentum variables are defined by
$M'_0=M_0 ({\bf k}_\perp\to {\bf k}'_\perp)$
with ${\bf k'}_\perp = {\bf k}_\perp + (1-x){\bf q}_{1\perp}$.
We confirmed numerically that Eq.~(\ref{Eq8}) exactly coincides with the manifestly covariant result given
by Eq.~(\ref{Eq6}). This verifies that the LF result obtained from the $q^+_1=0$ frame is immune to
the LF zero-mode contribution, which could have been the additional contribution right at
$p_1^+ = p_2^+ = 0$ if it exists. The LF zero mode involves the nonvalence wave function vertex discussed in our previous works~\cite{CRJ17,TWPS}. 
The Lorentz invariance of the TFF is complete in this work without any 
issue from the LF zero mode.

Since the amplitude of Fig.~\ref{fig2}(b) 
gives the same numerical values as that of Fig.~\ref{fig2}(a),
we obtain the total result  as 
$I^{m_Q}_{\rm tot}= 2 [I^{m_Q}_{(a)}]^{\rm Cov}= 2[I^{m_Q}_{(a)}]^{\rm LF}$.

\section{Application of the Light-Front Quark Model}
\label{sec:III}
In the standard LFQM~\cite{CJ_PLB,CJ_99,PiGam16,CJ_DA,CJBc,Jaus90,CCP,Choi07} approach, the
wave function of a ground state pseudoscalar meson as a $q\bar{q}$ bound state is given by
\be\label{QM1}
\Psi_{\lam{\bar\lam}}(x,{\bf k}_{\perp})
={\phi_R(x,{\bf k}_{\perp})\cal R}_{\lam{\bar\lam}}(x,{\bf k}_{\perp}),
\ee
where $\phi_R$ is the radial wave function and  ${\cal R}_{\lam{\bar\lam}}$
is the spin-orbit wave function
with the helicity $\lam~({\bar\lam})$ of a quark~(antiquark).

For the equal quark and antiquark mass $m_Q=m_{\bar Q}$,
the Gaussian wave function $\phi_R$ is given by
\be\label{QM2}
\phi_R(x,{\bf k}_{\perp})=
\frac{4\pi^{3/4}}{\beta^{3/2}}
\sqrt{\frac{M_0}{4 x (1-x)}} e^{m^2_Q/2\beta^2} e^{-M^2_0/8\beta^2},
\ee
where $\partial k_z/\partial x = M_0/4x(1-x)$ is the Jacobian of the variable transformation
$\{x,{\bf k}_\perp\}\to {\vec k}=({\bf k}_\perp, k_z)$
and $\beta$ is the variational parameter
fixed by our previous analysis of meson mass spectra~\cite{CJ_PLB,CJ_99,CJBc}.
The covariant form of the spin-orbit wave function ${\cal R}_{\lam{\bar\lam}}$
is given by
\be\label{QM4}
{\cal R}_{\lam{\bar\lam}}
=\frac{\bar{u}_{\lam}(p_Q)\gamma_5 v_{{\bar\lam}}( p_{\bar Q})}
{\sqrt{2}M_0},
\ee
and it satisfies
$\sum_{\lam{\bar\lam}}{\cal R}_{\lam{\bar\lam}}^{\dagger}{\cal R}_{\lam{\bar\lam}}=1$.
Thus, the normalization of our wave function is given by
\be\label{QM6}
\int^1_0 dx \int\frac{d^2{\bf k}_\perp}{16\pi^3}
|\phi_R(x,{\bf k}_{\perp})|^2 = 1.
\ee

In our previous analysis of the twist-2 and twist-3 DAs of
pseudoscalar and vector mesons~\cite{TWV,TWPS,TWPS17} and the pion electromagnetic form factor~\cite{TWPS},
we have shown that standard LF (SLF) results of the LFQM are obtained by
the replacement of the LF vertex function $\chi$ in the BS model with the Gaussian wave function
$\phi_R$ as follows  [see, e.g., Eq. (35) in~\cite{TWPS}] 
\be\label{QM7}
 \sqrt{2N_c} \frac{ \chi(x,{\bf k}_\perp) } {1-x}
 \to \frac{\phi_R (x,{\bf k}_\perp) }
 {\sqrt{{\bf k}^2_\perp + m_Q^2}}, \; M \to M_0,
 \ee
where $M\to M_0$ implies that the physical mass $M$ included in the integrand of BS
amplitude (except $M$ in the vertex function $\chi$) has to be replaced with the invariant
mass $M_0$ since the SLF results of the LFQM
are obtained from the requirement of all constituents being on their respective mass shell.
The correspondence in Eq.~(\ref{QM7}) is valid again in this analysis of a ${\rm P} \to \gamma^*\gamma^*$ transition.

Applying the correspondence given by Eq.~(\ref{QM7}) to  
$[I^{m_Q}_{(a)}]^{\rm LF}$ in Eq.~(\ref{Eq8}) and including the contribution
from Fig.~{\ref{fig2}(b) as well, 
we obtain the full result of  
$[I^{m_Q}_{\rm tot}]^{\rm LFQM}\equiv I^{m_Q}_{\rm QM}$ in our LFQM as follows:
\be\label{QM8b}
I^{m_Q}_{\rm QM} = \frac{\sqrt{2 N_c}}{4\pi^3}\int^{1}_0
 \frac{dx}{x} \int d^2{\bf k}_\perp
  \frac{m_Q}{(Q^2_2 + M^{\prime 2}_0)}
 \frac{\phi_R(x,{\bf k}_{\perp})}{\sqrt{{\bf k}^2_\perp + m^2_Q}}.
\ee
For $(\eta,\eta')\to\gamma^*\gamma^*$ transitions, 
making use of the $\eta-\eta'$ mixing scheme,
the flavor assignment of $\eta$ and $\eta'$ mesons in the quark-flavor basis $\eta_q=(u\bar{u}+d\bar{d})/\sqrt{2}$ and
 $\eta_s=s\bar{s}$ is given by~\cite{FKS}
  \be\label{Eq7a}
 \left( \begin{array}{cc}
 \eta\\
 \eta'
 \end{array}\,\right)
 =\left( \begin{array}{cc}
 \cos\phi\;\; -\sin\phi\\
 \sin\phi\;\;\;\;\;\cos\phi
 \end{array}\,\right)\left( \begin{array}{c}
 \eta_q\\
 \eta_s
 \end{array}\,\right).
 \ee

Using this mixing scheme and including the electric charge factors, we obtain the transition form factors
$F_{{\rm P}\gamma^*}(Q^2_1,Q^2_2)$ for ${\rm P}\to\gamma^*\gamma^* ~({\rm P}=\pi^0, \eta, \eta')$ transitions
as follows
\bea\label{Eq5}
F_{\pi\gamma^*}(Q^2_1, Q^2_2) &=& \frac{(e^2_u - e^2_d)}{\sqrt{2}} I^{m_{u(d)}}_{\rm QM},
\nonumber\\
F_{\eta\gamma^*} (Q^2_1,Q^2_2) &=& \cos\phi\; \frac{(e^2_u + e^2_d)}{\sqrt{2}} I^{m_{u(d)}}_{\rm QM}
- \sin\phi\; e^2_s  I^{m_s}_{\rm QM},
\nonumber\\
F_{\eta'\gamma^*} (Q^2_1,Q^2_2) &=& \sin\phi\;\frac{(e^2_u + e^2_d)}{\sqrt{2}} I^{m_{u(d)}}_{\rm QM}
+ \cos\phi\; e^2_s  I^{m_s}_{\rm QM}.
\eea
While the quadratic (linear) Gell-Mann-Okubo mass formula prefers
$\phi\simeq 44.7^\circ$ ($\phi\simeq 31.7^\circ$)~\cite{PDG18},
the KLOE Collaboration~\cite{KLOE} extracted the pseudoscalar mixing
angle $\phi$
by measuring the ratio ${\rm BR}(\phi\to\eta'\gamma)/{\rm BR}(\phi\to\eta\gamma)$.
The measured values are $\phi=(39.7\pm 0.7)^{\circ}$ and
$(41.5\pm 0.3_{\rm stat}\pm 0.7_{\rm syst}\pm 0.6_{\rm th})^{\circ}$
with and without the gluonium content for $\eta'$, respectively.
In this work, however, we use $\phi=37^\circ\pm 5^\circ$ to check the sensitivity of our LFQM.

For a sufficiently high spacelike momentum transfer $(Q^2_1, Q^2_2)$
region, our LFQM result for $F_{\pi\gamma^*}(Q^2_1,Q^2_2)$   can be approximated
in the leading order (LO) as follows:
\be\label{QM10}
F^{\rm LO}_{\pi\gamma^*}(Q^2_1,Q^2_2)\simeq C_\pi
\int^{1}_0 dx\; \frac{\phi_{2;\pi} (x)}{(1-x)Q^2_1 + xQ^2_2},
\ee
where $C_\pi=(\sqrt{2} /3)f_\pi$, with 
$f_\pi$ the pseudoscalar meson decay constant and  $\phi_{2;\pi}(x)$
is the twist-2 pion distribution amplitude (DA) in our LFQM given by~\cite{TWV,TWPS,TWPS17}
\bea\label{2DA}
\phi_{2;\pi} (x) &=& \frac {\sqrt{2N_c}} {f_\pi 8\pi^3} \int d^2{\bf k}_\perp
\frac{\phi_{R}(x, {\bf k}_\perp)}{\sqrt{ {\bf k}^2_\perp + m_Q^2} }{m_Q}.
\eea
Our result for $\phi_{2;\pi} (x)$ can be found in Ref.~\cite{CJ_DA}. 
As one can see from Eq.~(\ref{QM10}), while the singly virtual TFF 
$F_{\pi\gamma^*}(Q^2, 0)$ above some intermediate values of momentum transfer
is known to be quite sensitive to the shape of DA,
the doubly virtual TFF is not sensitive to the shape of DA since the amplitude $T_H = 1/( (1-x)Q^2_1 + xQ^2_2)$
is finite at the end points of $x$, i.e. $x=0, 1$.

We note that the pQCD LO result for $F_{\pi\gamma^*}(Q^2_1,Q^2_2)$ can be obtained from
replacing $\phi_{2;\pi} (x)$ in Eq.~(\ref{QM10}) with 
the asymptotic form $\phi^{asy}_{2;\pi} (x) = 6 x(1-x)$~\cite{BL80}. 
Taking the same asymptotic form $6 x(1-x)$ for the quark DAs,
the pQCD LO results for $(\eta, \eta')$ TFFs 
can also be obtained by 
replacing the factor $C_\pi$ in Eq.~(\ref{QM10})
with $C_\eta=(5\sqrt{2} /9)f_{\eta_q}\cos\phi - (2 /9)f_{\eta_s}\sin\phi$ for $F_{\eta\gamma^*} (Q^2_1,Q^2_2)$
and with 
$C_{\eta'}=(5\sqrt{2} /9)f_{\eta_q}\sin\phi + (2 /9)f_{\eta_s}\cos\phi$ for $F_{\eta'\gamma^*} (Q^2_1,Q^2_2)$,
where $f_{\eta_q}$ and $f_{\eta_s}$ are the weak decay constants for the $|\eta_q\ra$ 
and $|\eta_s\ra$ states, respectively.
For this transition to two highly off-shell photons, the pQCD expression for the next-to-leading order (NLO) component
can be found in Ref.~\cite{Braaten83}.

\section{Numerical Results}
\label{sec:IV}
\begin{table}[t]
\caption{The constituent quark masses $m_Q(Q=u(d), s)$ (in GeV) and the Gaussian parameters
$\beta_{Q{\bar Q}}$ (in GeV) for the linear confining potentials
obtained from the variational principle in our LFQM~\cite{CJ_PLB,CJ_99,CJ_DA}.}
\label{t1}
\renewcommand{\tabcolsep}{1.5pc} 
\begin{tabular}{cccc} \hline\hline
 $m_{u(d)}$ & $m_s$ & $\beta_{Q{\bar Q}}$ & $\beta_{s{\bar s}}$ \\
\hline
 0.22~ & 0.45~ & 0.3659~ & 0.4128~  \\
\hline\hline
\end{tabular}
\end{table}
In our numerical calculations within the standard LFQM, we use the model parameters
(i.e. constituent quark masses $m_Q$ and Gaussian parameters $\beta_{Q{\bar Q}}$) for the
linear confining potentials given in Table~\ref{t1}, which were obtained from the calculation of meson mass spectra using 
the variational principle in our LFQM~\cite{CJ_PLB,CJ_99,CJ_DA}. The analysis for singly virtual TFFs 
$F_{{\rm P}\gamma}(Q^2,0)$ can be found in our previous work~\cite{CRJ17}.

\begin{figure}
\vspace{0.5cm}
\begin{center}
\includegraphics[height=6cm, width=7cm]{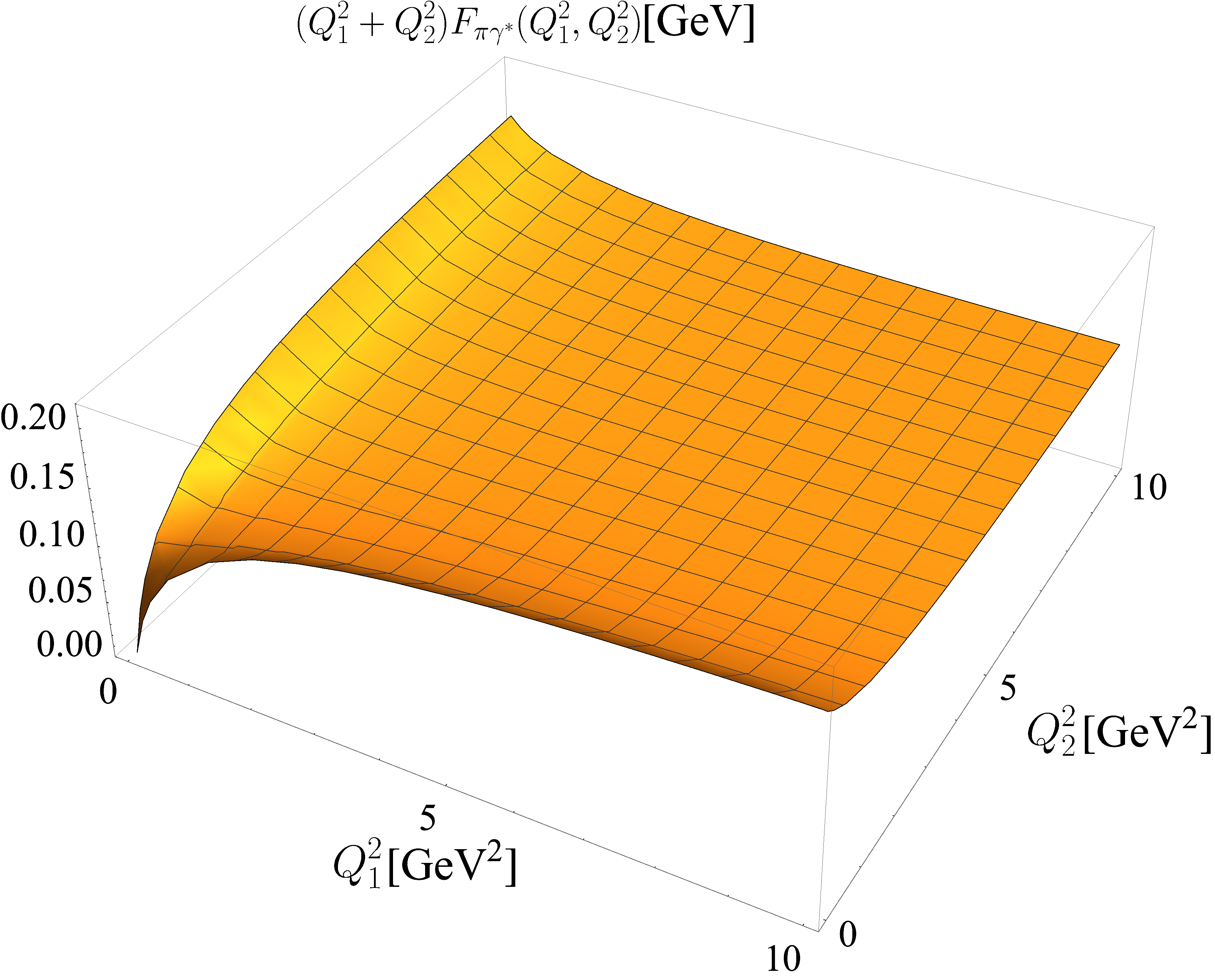}
\\
\includegraphics[height=6cm, width=7cm]{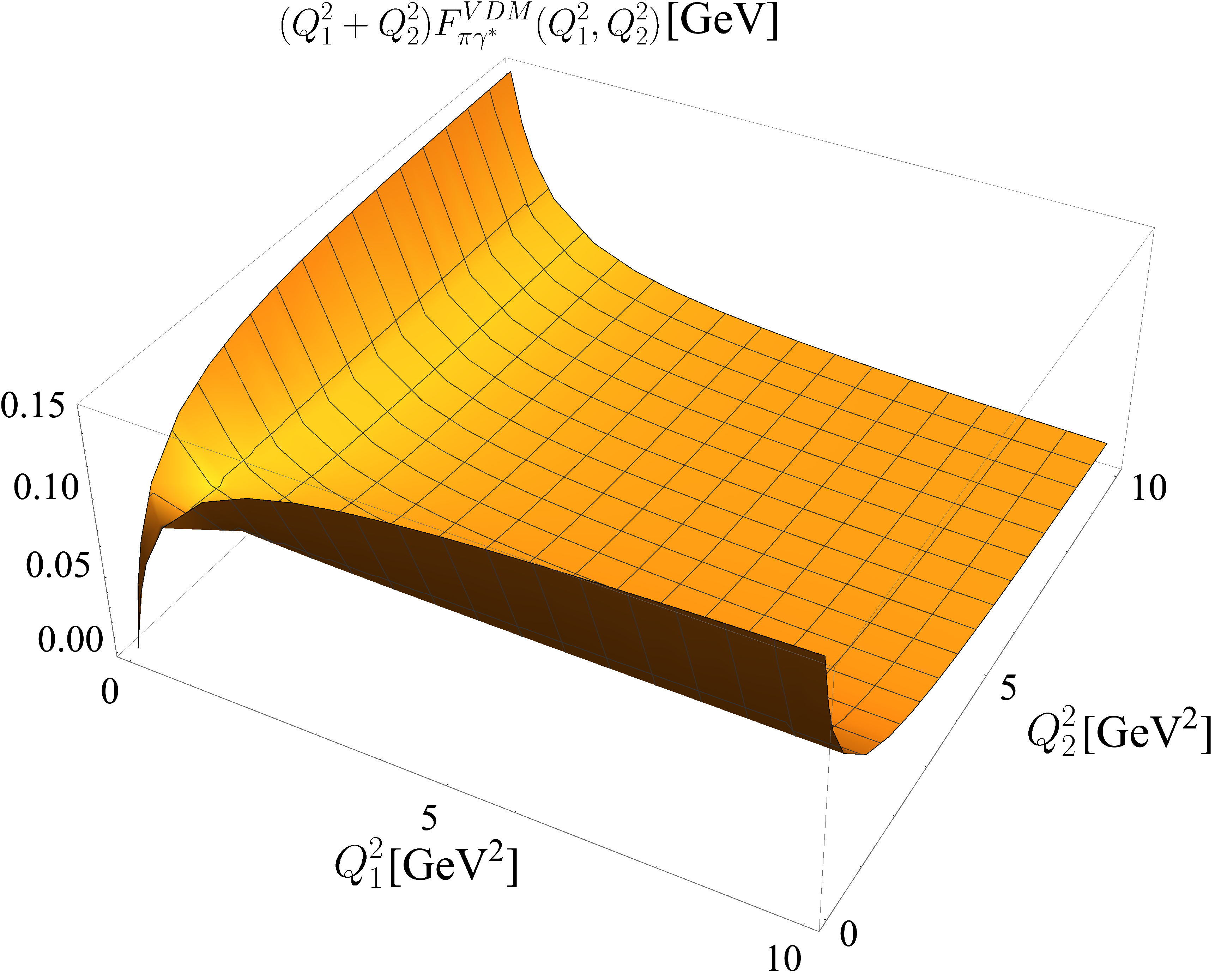}
\caption{\label{fig3} The three-dimensional plots for $(Q^2_1 + Q^2_2) F_{\pi\gamma^*}(Q^2_1,Q^2_2)$ obtained 
from Eq.~(\ref{Eq5}) (upper panel) compared with the VMD result (lower panel) for the range 
of $0< (Q^2_1, Q^2_2)<10$ GeV$^2$. }
\end{center}
\end{figure}
In Fig.~\ref{fig3}, we show the three-dimensional plots for $(Q^2_1 + Q^2_2) F_{\pi\gamma^*}(Q^2_1,Q^2_2)$ 
for the $0< (Q^2_1, Q^2_2)<10$ GeV$^2$ range obtained from Eq.~(\ref{Eq5}) and compare our LFQM result (upper panel)
with the result from the VMD model (lower panel), which is given by~\cite{BABAR18}
\be
F^{\rm VMD}_{{\rm P}\gamma^*}(Q^2_1,Q^2_2) 
= \frac{ F_{{\rm P}\gamma}(0,0)}{(1+ Q^2_1/\Lambda^2_{\rm P})(1+ Q^2_2/\Lambda^2_{\rm P})},
\ee
where we take $\Lambda_{\rm P}=775$ MeV corresponding to the $\rho$-pole and the central value of the 
experimental data~\cite{PDG18}, $F^{\rm Exp.}_{\pi\gamma}(0,0)=0.272(3)$ GeV$^{-1}$
for $F_{\pi\gamma}(0,0)$. 
As we discussed before, while our LFQM result for doubly virtual TFF behaves as
$F_{\pi\gamma^*}(Q^2_1, Q^2_2)\sim 1/(Q^2_1 + Q^2_2)$ as $(Q^2_1, Q^2_2)\to\infty$, which is  consistent with the pQCD prediction, 
the result of the VMD model behaves as $F^{\rm VMD}_{\pi\gamma^*}(Q^2_1, Q^2_2)\sim 1/(Q^2_1 Q^2_2)$.  
On the other hand, for the singly virtual TFF such
as $F_{\pi\gamma^*}(Q^2_1=Q^2, 0)$  or $F_{\pi\gamma^*}(0, Q^2_2=Q^2)$, the two models show the same scaling behavior 
$Q^2 F_{\pi\gamma^*}(Q^2,0)\to{\rm constant}$. 
One can also see from Fig.~\ref{fig3} that our LFQM result for the TFF is in general larger in the asymmetric limit (e.g., $Q^2_1=Q^2, Q^2_2=0$)
than in the symmetric limit (i.e., $Q^2_1=Q^2_2$), which persists up to an asymptotically large momentum transfer region. The same observation
was made in Ref.~\cite{Weil}.

\begin{figure}
\includegraphics[height=8cm, width=7cm]{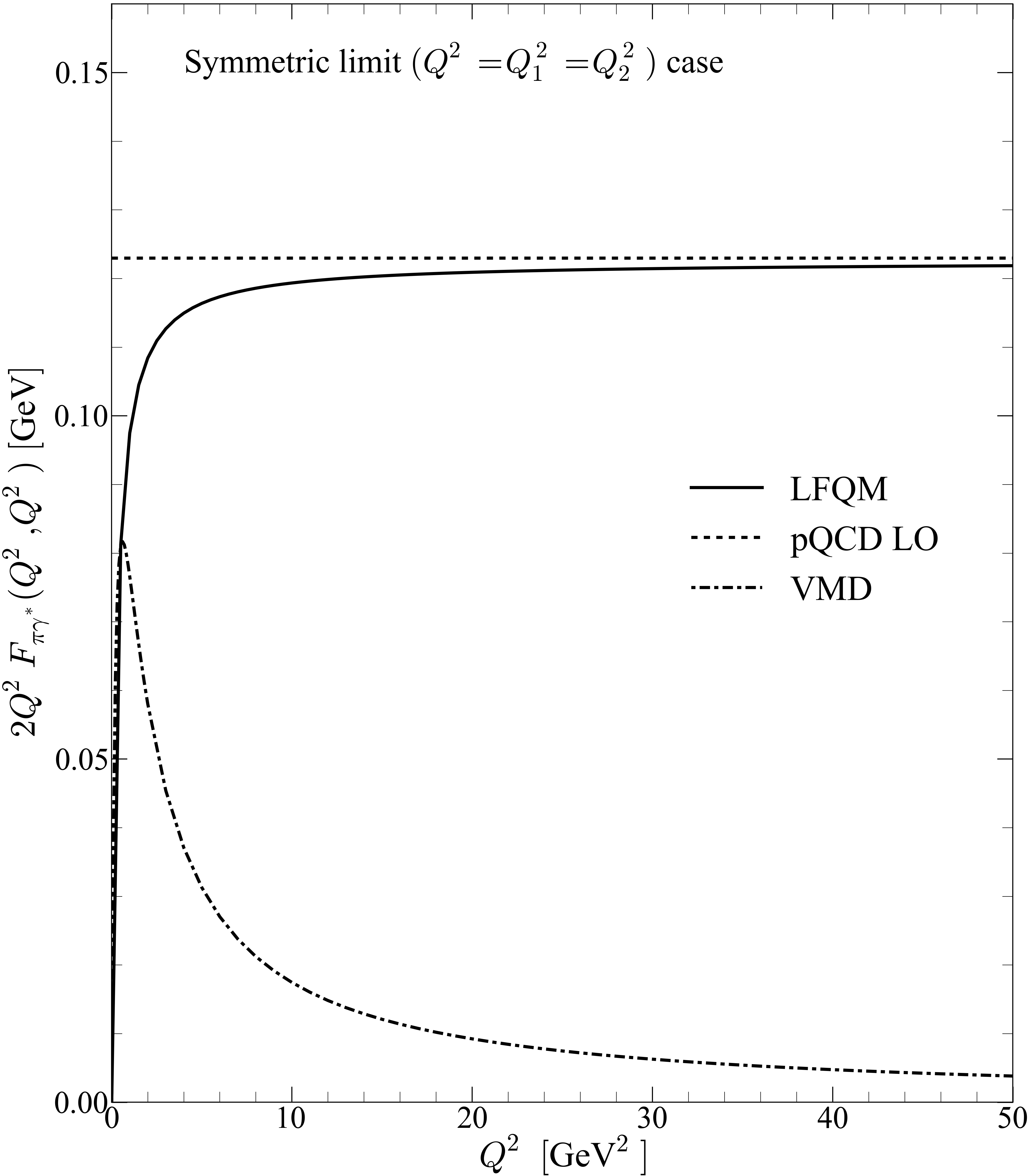}
\caption{\label{fig4} The two-dimensional plot for $2 Q^2 F_{\pi\gamma^*}(Q^2,Q^2)$  in the
symmetric limit ($Q^2=Q^2_1=Q^2_2)$ for the $0<Q^2<50$ GeV$^2$ region compared with the pQCD LO
and the VMD model predictions.}
\end{figure}

In Fig.~\ref{fig4}, we show the two-dimensional plot for $2 Q^2 F_{\pi\gamma^*}(Q^2,Q^2)$  in the
symmetric limit ($Q^2=Q^2_1=Q^2_2)$ for the $0<Q^2<50$ GeV$^2$ region compared with the pQCD LO
and the VMD model predictions. In this symmetric limit case, the different behavior of $Q^2 F_{\pi\gamma^*}(Q^2,Q^2)$ 
between our LFQM result (solid line) $Q^2 F_{\pi\gamma^*}(Q^2,Q^2)\to{\rm constant}$ and
the VMD result (dotted-dashed line)  $Q^2F_{\pi\gamma^*}(Q^2,Q^2)\to 1/Q^2$ can be clearly seen
as $Q^2\to\infty$.  Comparing our LFQM result and the pQCD LO (dashed line) prediction, while
the NLO contribution is still greater than 10$\%$ for the  $Q^2\leq 2$ GeV$^2$ region, the NLO contribution
becomes less than 5$\%$ for the $Q^2\geq 6$ GeV$^2$ region. 

\begin{figure}
\vspace{0.5cm}
\begin{center}
\includegraphics[height=6cm, width=7cm]{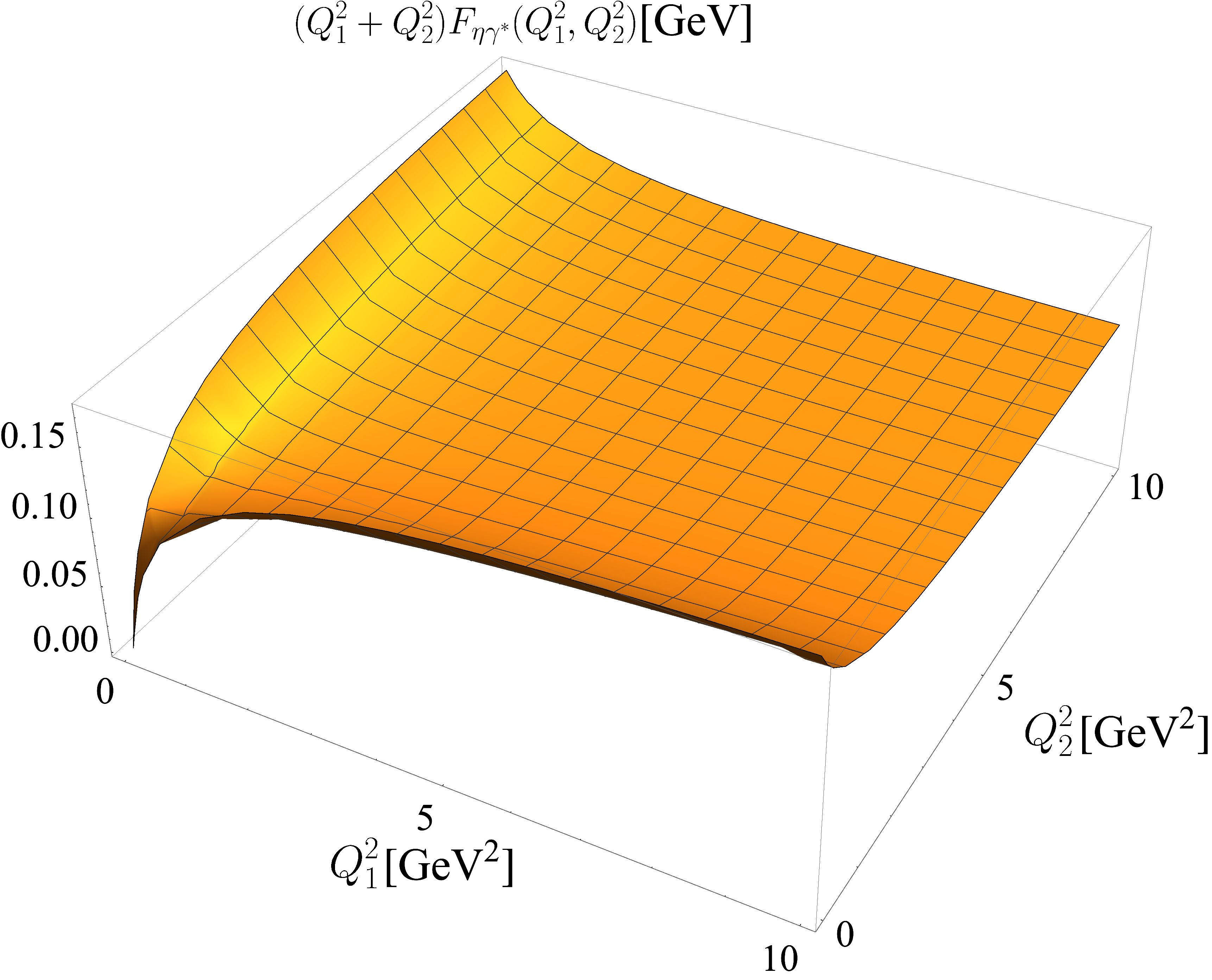}
\\
\includegraphics[height=6cm, width=7cm]{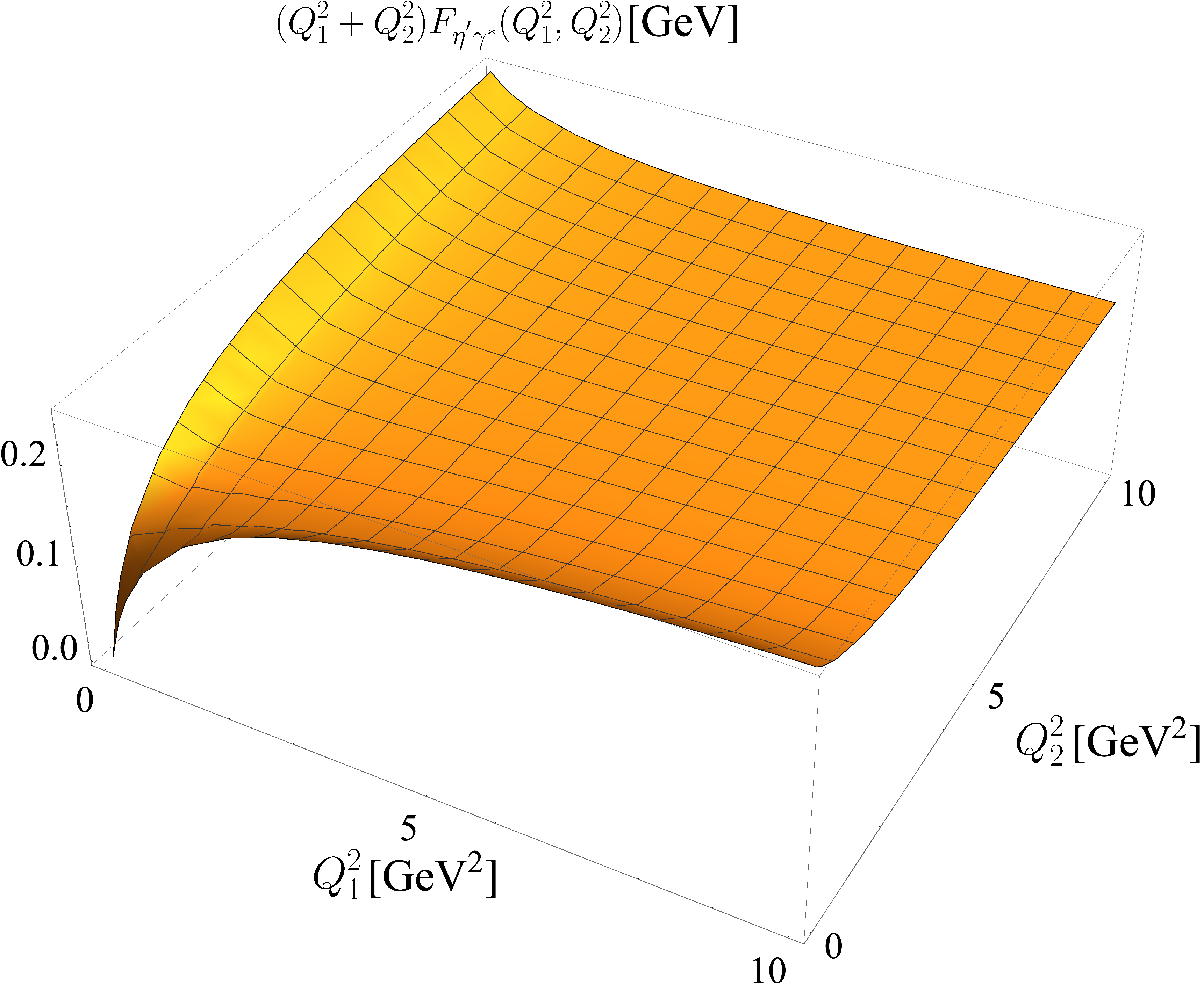}
\caption{\label{fig5} The three-dimensional plots for $(Q^2_1 + Q^2_2) F_{\eta\gamma^*}(Q^2_1,Q^2_2)$ (upper panel)
and $(Q^2_1 + Q^2_2) F_{\eta'\gamma^*}(Q^2_1,Q^2_2)$ (lower panel) obtained 
from Eq.~(\ref{Eq5}) with $\phi=37^\circ$ for the range of $0< (Q^2_1, Q^2_2)<10 $ GeV$^2$. }
\end{center}
\end{figure}
In Fig.~\ref{fig5}, we show the three-dimensional plots for $(Q^2_1 + Q^2_2) F_{\eta\gamma^*}(Q^2_1,Q^2_2)$ (upper panel)
and $(Q^2_1 + Q^2_2) F_{\eta'\gamma^*}(Q^2_1,Q^2_2)$ (lower panel) obtained 
from Eq.~(\ref{Eq5}) with $\phi=37^\circ$ for the range of $0< (Q^2_1, Q^2_2)<10 $ GeV$^2$.
As one can see from Figs.~\ref{fig3} and~\ref{fig4}, all three TFFs $F_{(\pi, \eta,
\eta')\gamma^*}(Q^2_1,Q^2_2)$ obtained from our LFQM show the same scaling behavior as the pQCD predicted.

\begin{table*}[t]
\caption{The  transition form factors $F_{(\pi, \eta,\eta')\gamma^*}(Q^2_1,Q^2_2)$ (in units of $10^3$ GeV$^{-1}$) 
for some ($Q^2_1,Q^2_2$) values (in units of GeV$^2$) compared with the experimental data~\cite{BABAR18} 
for $F^{\rm Exp.}_{\eta'\gamma^*}$.}
\label{t2}
\renewcommand{\tabcolsep}{1.5pc} 
\begin{tabular}{ccccc} \hline\hline
 $(Q^2_1,Q^2_2)$ & $F_{\pi\gamma^*}$ & $F_{\eta\gamma^*}$ & $F_{\eta'\gamma^*}$ & $F^{\rm Exp.}_{\eta'\gamma^*}$ \\
\hline
 (6.48,6.48) & 9.08~ & $8.48^{+1.18}_{-1.24}$~ & $13.91^{+0.69}_{-0.79}$~ & $14.32^{+1.95}_{-1.89}\pm 0.83\pm 0.14$~ \\
 \hline
 (16.85,16.85) & 3.58~ & $3.29^{+0.47}_{-0.50}$~ & $5.55^{+0.27}_{-0.31}$~ & $5.35^{+1.71}_{-2.15}\pm 0.31\pm 0.42$~ \\
  \hline
 (14.83,4.27) & 6.76~ & $6.33^{+0.87}_{-0.92}$~ & $10.32^{+0.51}_{-0.59}$~ & $8.24^{+1.16}_{-1.13}\pm 0.48\pm 0.65$~ \\
  \hline
 (38.11,14.95) & 2.40~ & $2.21^{+0.32}_{-0.33}$~ & $3.71^{+0.18}_{-0.21}$~ & $6.07^{+1.09}_{-1.07}\pm 0.35\pm 1.21$~ \\
  \hline
 (45.63,45.63) & 1.33~ & $1.22^{+0.18}_{-0.19}$~ & $2.08^{+0.10}_{-0.11}$~ & $8.71^{+3.96}_{-4.02}\pm 0.50\pm 1.04$~ \\
\hline\hline
\end{tabular}
\end{table*}
In Table~\ref{t2}, we summarize our LFQM results for the transition form factors $F_{(\pi, \eta,\eta')\gamma^*}(Q^2_1,Q^2_2)$ (in units of $10^3$ GeV$^{-1}$) 
for some ($Q^2_1,Q^2_2$) values (in units of GeV$^2$) compared with the experimental data~\cite{BABAR18} 
for $F^{\rm Exp.}_{\eta'\gamma^*}$ with the statistical, systematic, and model uncertainties.
We note that the error estimates for $F_{(\eta,\eta')\gamma^*}(Q^2_1,Q^2_2)$ in our LFQM results come from
the choice of $\eta-\eta'$ mixing angle $\phi=(37\pm 5)^\circ$.
We note for $F_{\eta'\gamma^*}(Q^2_1,Q^2_2)$ that our LFQM result and the experimental data 
are compatible with each other and the agreement between the two appears fairly reasonable
within a rather large uncertainty of data.

\begin{figure*}
\vspace{0.5cm}
\begin{center}
\includegraphics[height=6cm, width=8cm]{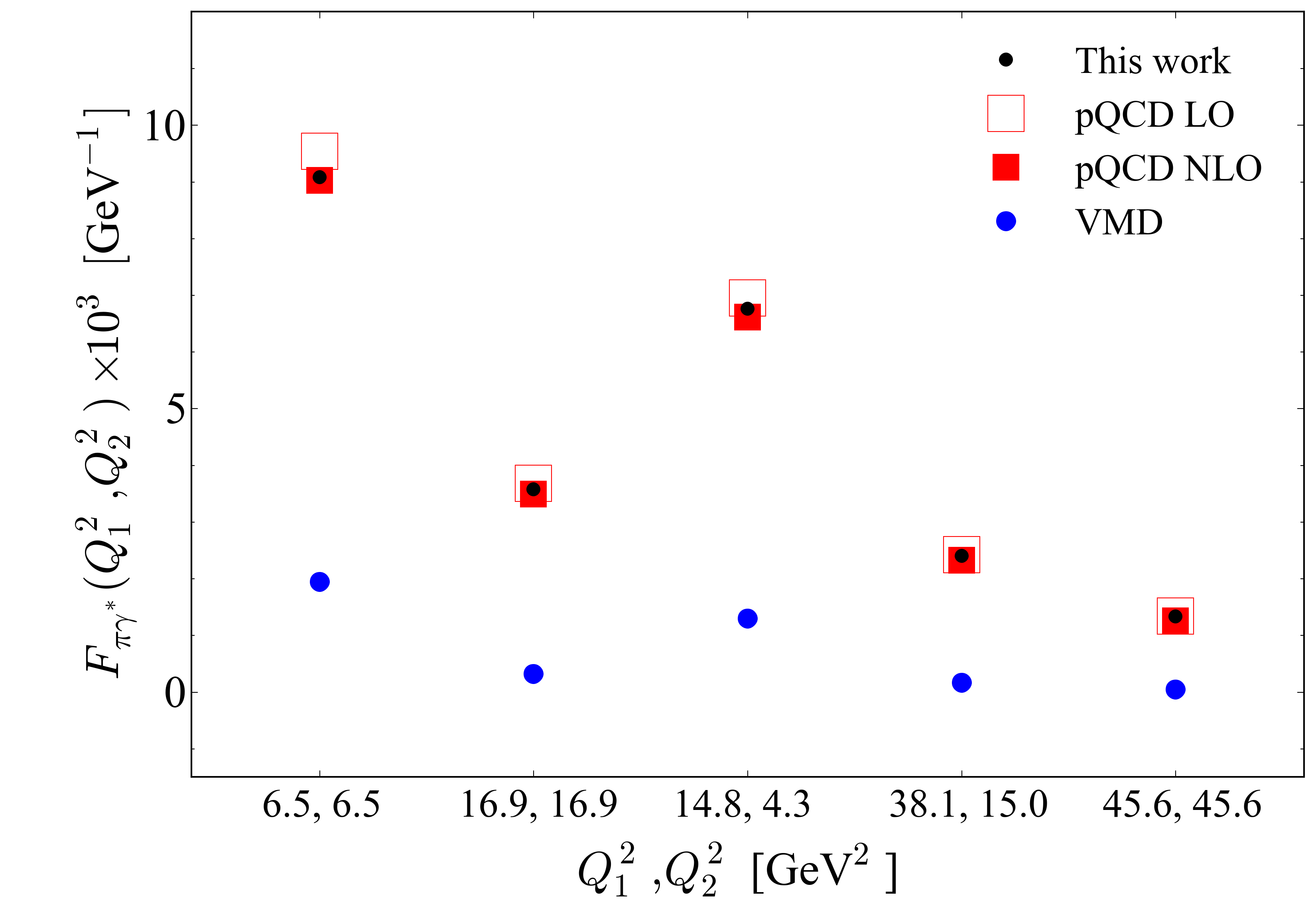}
\includegraphics[height=6cm, width=8cm]{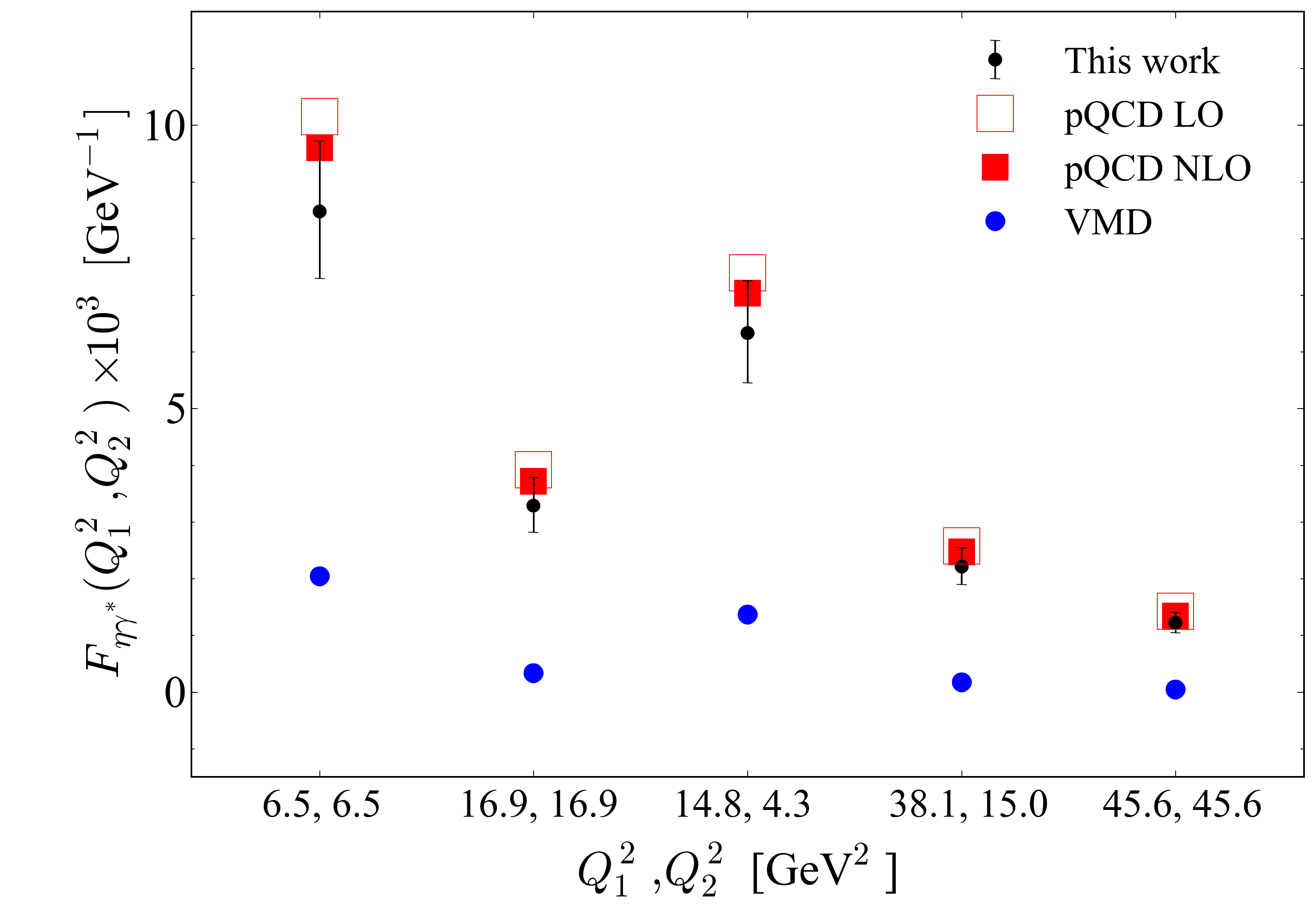}
\\
\includegraphics[height=6cm, width=8cm]{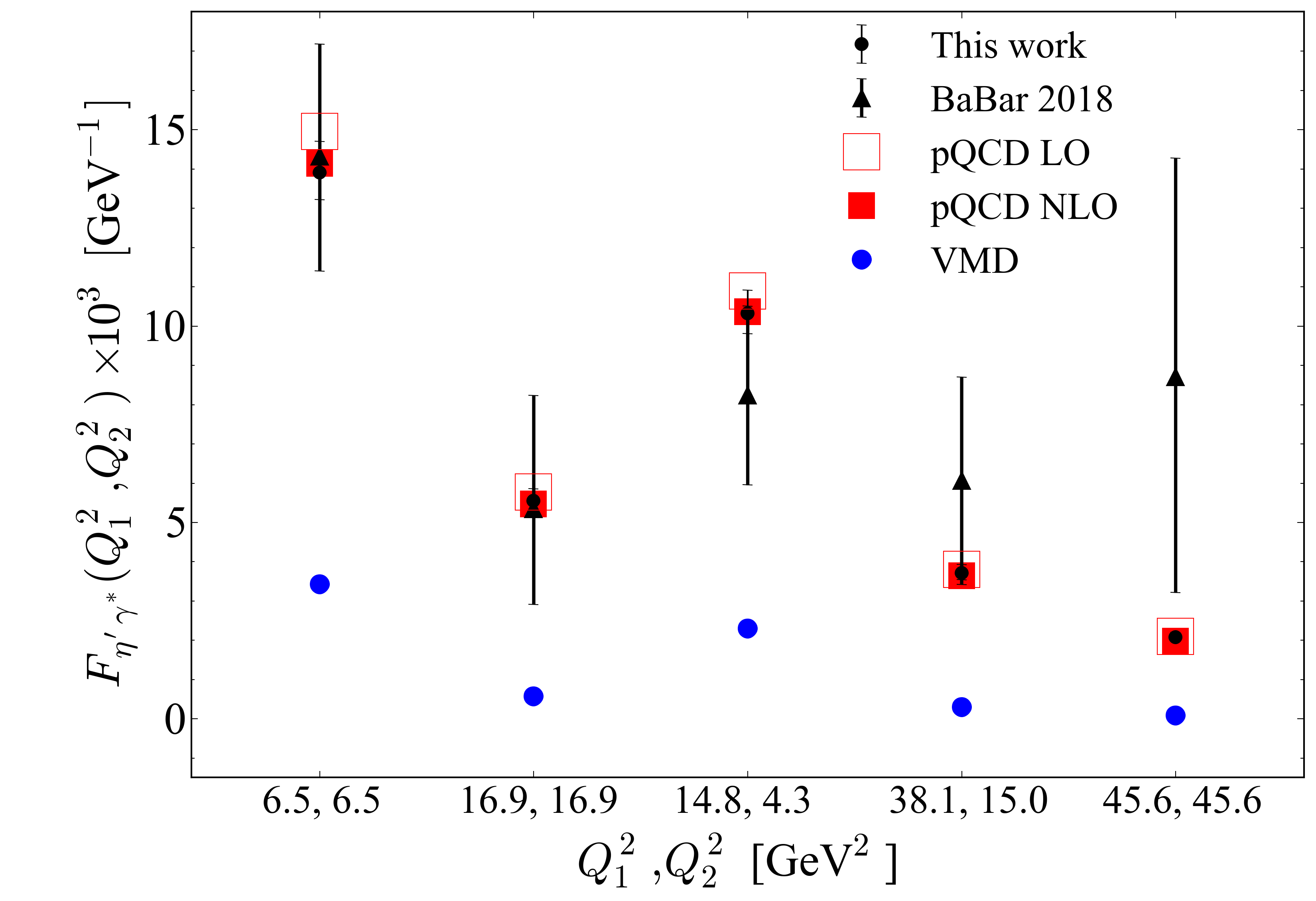}
\caption{\label{fig6} Our LFQM results for $F_{(\pi,\eta,\eta')\gamma^*}(Q^2_1,Q^2_2)$ (black circles) compared with the
pQCD LO (open squares) and NLO(filled squares) predictions; the VMD predictions (blue circles); and the 
experimental data~\cite{BABAR18}~(triangles, with error bars
including the statistical, systematic, and model uncertainties).}
\end{center}
\end{figure*}
In Fig.~\ref{fig6}, we show our LFQM results for $F_{(\pi,\eta,\eta')\gamma^*}(Q^2_1,Q^2_2)$ (black circles) compared with the pQCD
LO (open squares) and NLO(filled squares) predictions~\cite{Braaten83}, VMD predictions (blue circles), and the experimental data~\cite{BABAR18} (triangles) for
$F_{\eta'\gamma^*}(Q^2_1,Q^2_2)$. We note that the error bars for $F^{\rm Exp.}_{\eta'\gamma^*}(Q^2_1,Q^2_2)$  include the
statistical, systematic, and model uncertainties. As one can see from Fig.~\ref{fig5}, our LFQM results 
for $F_{(\pi,\eta,\eta')\gamma^*}(Q^2_1,Q^2_2)$ show the same behavior as the 
 pQCD predictions. However, our LFQM predictions are quite different from the VMD model
predictions since the two models have different power behaviors of $(Q^2_1, Q^2_2)$ as we discussed before. 
While the data for $F_{\eta'\gamma^*}(Q^2_1,Q^2_2)$ measured from  $BABAR$~\cite{BABAR18}  
agree with the pQCD and our LFQM predictions, they show a clear disagreement with the VMD model predictions.

\section{Summary and Discussion}
\label{sec:V}
We presented the doubly  virtual TFFs $F_{{\rm P}\gamma^*}(Q^2_1,Q^2_2)$ for the 
${\rm P}\to\gamma^*\gamma^* \;({\rm P}=\pi^0,\eta,\eta')$ transitions  in the standard LF (SLF) approach within the 
phenomenologically accessible realistic LFQM~\cite{CJ_PLB,CJ_99, PiGam16,CJ_DA,CJBc}.  
Performing a LF calculation in the covariant BS model as the first illustration, 
we used the $q^+_1=0$ frame with $q^2_1=-{\bf q}^2_{1\perp}=-Q^2_1$, and we found that both LF and manifestly covariant calculations produced
exactly the same results for $F_{{\rm P}\gamma^*}(Q^2_1,Q^2_2)$. This assured the absence of the LF zero mode in the doubly virtual TFFs 
as expected~\cite{CRJ17}. 

We then mapped the exactly solvable manifestly covariant BS model to the standard LFQM following the same correspondence 
relation given by Eq.~(\ref{QM7}) between 
the two models that we found in our previous analysis of two-point and three-point functions for the pseudoscalar and vector 
mesons~\cite{TWV,TWPS}. This allowed us to apply the more phenomenologically accessible 
Gaussian wave function provided by the LFQM analysis 
of meson mass spectra~\cite{CJ_PLB,CJ_99, PiGam16,CJ_DA,CJBc} to the analysis of  the doubly virtual $F_{{\rm P}\gamma^*}(Q^2_1,Q^2_2)$.
For the $(\eta,\eta')\to\gamma^*\gamma^*$ transitions, we used the $\eta-\eta'$
mixing angle $\phi$ in the quark-flavor basis varying the $\phi$ values in the range of $\phi=(37\pm 5)^\circ$ to check the sensitivity of our LFQM.

For the numerical analyses of $F_{{\rm P}\gamma^*}(Q^2_1,Q^2_2)$, 
we compared our LFQM results with the available experimental data and the other theoretical model predictions
such as the pQCD~\cite{Braaten83} and VMD results.
While our LFQM result for the doubly virtual TFF behaves as
$F_{{\rm P}\gamma^*}(Q^2_1, Q^2_2)\sim 1/(Q^2_1 + Q^2_2)$ as $(Q^2_1, Q^2_2)\to\infty$, which is consistent with the pQCD prediction, 
the result of the VMD model behaves as $F^{\rm VMD}_{{\rm P}\gamma^*}(Q^2_1, Q^2_2)\sim 1/(Q^2_1 Q^2_2)$.
 Our LFQM prediction for 
$F_{\eta'\gamma^*}(Q^2_1,Q^2_2)$ showed a reasonable agreement with the very recent experimental data  obtained from the $BABAR$ collaboration
for the ranges of $2< (Q^2_1, Q^2_2) <60$ GeV$^2$.

\acknowledgments
H.-M.C. was supported by the National Research Foundation of Korea~(NRF)
(Grant No. NRF-2017R1D1A1B03033129). H.-Y. R. was supported by the NRF grant funded by
the Korean government~(MSIP)~(Grant No. 2015R1A2A2A01004238).
C.-R. J. was supported in part by the U.S. Department of Energy~
(Grant No. DE-FG02-03ER41260).


\begin{thebibliography}{99}



\bibitem{JN} F. Jegerlehner and A. Nyffeler, Phys. Rep. {\bf 477}, 1 (2009).

\bibitem{Ny2016} A. Nyffeler, Phys. Rev. D {\bf 94}, 053006 (2016).

\bibitem{Lattice16} A. G\'erardin, H. Meyer, and A. Nyffeler,
\Journal{\PRD}{94}{074507}{2016}.

\bibitem{CELLO91}  H.-J. Behrend {\em et al.} (CELLO Collaboration),
Z. Phys. C {\bf 49}, 401 (1991).

\bibitem{CLEO98} J. Gronberg {\em et al.} (CLEO Collaboration),
\Journal{\PRD}{57}{33}{1998}.

\bibitem{BES15_Pi} A. Denig (BESIII Collaboration), Nucl. Part. Phys. Proc. {\bf 260}, 79 (2015).

\bibitem{NA60} R. Arnaldi {\em et al.} (NA60 Collaboration),
\Journal{\PLB}{677}{260}{2009}.

\bibitem{NA60-17} C. Lazzeroni {\em et al.} (NA62 Collaboration), 
\Journal{\PLB}{768}{38}{2017}.

\bibitem{A22014} P. Aguar-Bartomom\'{e} {\em et al.} (A2 Collaboration), 
\Journal{\PRC}{89}{044608}{2014}.

\bibitem{A22011} H. Bergh\"{a}user {\em et al.} (A2 Collabortation), 
\Journal{\PLB}{701}{562}{2011}.

\bibitem{A2pi} P. Adlarson {\em et al.} (A2 Collaboration),  \Journal{\PRC}{95}{025202}{2017}.

\bibitem{BES15} M. Ablikim {\em et al.} (BESIII Collaboration), Phys. Rev. D {\bf 92}, 012001 (2015).

\bibitem{BABAR06} B. Aubert {\em et al.} ($BABAR$ Collaboration), Phys. Rev. D {\bf 74}, 012002 (2006). 

\bibitem{BABAR18} J. P. Lees {\em et al.} ($BABAR$ Collaboration), \Journal{\PRD}{98}{112002}{2018}. 

\bibitem{BL80} G.P. Lepage and S.J. Brodsky, \Journal{\PRD}{22}{2157}{1980}.

\bibitem{Braaten83} E. Braaten, \Journal{\PRD}{28}{524}{1983}.

\bibitem{VDM1} B.-I. Young, Phys. Rev.  {\bf 161},  1620 (1967).

\bibitem{VDM2} L. G. Landsberg, Phys. Rep.  {\bf 128},  301 (1985).

\bibitem{VDM3} A. Dorokhov, M. Ivanov, and S. Kovalenko, \Journal{\PLB}{677}{145}{2009}.

\bibitem{Weil} E. Weil, G. Eichmann, C. S. Fischer, and R. Williams, \Journal{\PRD}{96}{014021}{2017}.

\bibitem{CRJ17} H.-M. Choi, H. -Y. Ryu, and C.-R. Ji, \Journal{\PRD}{96}{056008}{2017}.

\bibitem{CJ_99} H.-M. Choi and C.-R. Ji, \Journal{\PRD}{59}{074015}{1999}.

\bibitem{CJ_DA} H.-M. Choi and C.-R. Ji, \Journal{\PRD}{75}{034019}{2007}.

\bibitem{PiGam16} H.-M. Choi and C.-R. Ji, Few-Body Syst. {\bf 57}, 497 (2016).

\bibitem{CJ_PLB} H.-M. Choi and C.-R. Ji, \Journal{\PLB}{460}{461}{1999}.


\bibitem{CJBc} H.-M. Choi and C.-R. Ji, \Journal{\PRD}{80}{054016}{2009}.

\bibitem{Zero1} M. Burkardt, \Journal{\PRD}{47}{4628}{1993}.

\bibitem{Zero2} S. J. Brodsky and D. S. Hwang, \Journal{\NPB}{543}{239}{1999}.

\bibitem{Zero3} J.P.B.C. de Melo, J.H.O. Sales, T. Frederico, and P.U. Sauer, \Journal{\NPA}{631}{574c}{1998}.

\bibitem{Zero4} H.-M. Choi and C.-R. Ji, \Journal{\PRD}{58}{071901(R)}{1998}.


\bibitem{BCJ02} B.L.G. Bakker, H.-M. Choi, and C.-R. Ji, \Journal{\PRD}{65}{116001}{2002}.

\bibitem{BCJ03} B.L.G. Bakker, H.-M. Choi, and C.-R. Ji, \Journal{\PRD}{67}{113007}{2003}.

\bibitem{TWV} H.-M. Choi and C.-R. Ji,
\Journal{\PRD}{89}{033011}{2014}.

\bibitem{TWPS} H.-M. Choi and C.-R. Ji,
\Journal{\PRD}{91}{014018}{2015}.

\bibitem{TWPS17} H.-M. Choi and C.-R. Ji,
\Journal{\PRD}{95}{056002}{2017}; Few-Body Syst. {\bf 58}, 31 (2017).

\bibitem{FKS}  T. Feldmann, P. Kroll, and B. Stech,
\Journal{\PRD}{58}{114006}{1998}.

\bibitem{Jaus90} W. Jaus, \Journal{\PRD}{41}{3394}{1990}.


\bibitem{CCP} P. L. Chung, F. Coester, and W. N. Polyzou,
\Journal{\PLB}{205}{545}{1988}.

\bibitem{Choi07} H.-M. Choi, \Journal{\PRD}{75}{073016}{2007}.

\bibitem{PDG18} M. Tanabashi {\em et al.} (Particle Data Group), 
\Journal{\PRD}{98}{030001}{2018}.

\bibitem{KLOE} F. Ambrosino {\em et al.} (KLOE Collaboration),
\Journal{\PLB}{648}{267}{2007}.




































































































\end{thebibliography}
\end{document}